\documentclass[
    ,final            
  ]
  {aipproc}

\usepackage{microtype}
\layoutstyle{8x11single}

\def\fm {\,\rm{\hbox{fm}}}
\def\MeV {\,\rm{\hbox{MeV}}}

\def\solution#1  {}

\usepackage{amsmath} 
\usepackage{dsfont}  
\usepackage{bm}      

\def\beq{\begin{equation}}
\def\eeq{\end{equation}}
\def\beqs#1\eeqs{\beq\begin{split} #1 \end{split}\eeq}

\def\comment#1{}

\begin{document}

\title{Pion electric polarizability from lattice QCD}

\classification{14.40.Be,11.15.Ha,12.38.Gc}
\keywords      {lattice QCD, polarizability}

\author{Andrei Alexandru}{}
\author{Michael Lujan}{}
\author{Walter Freeman}{}
\author{Frank Lee}{
  address={The George Washington University, 725 21st St. NW, Washington DC, 20052}
}

\begin{abstract}
 Electromagnetic polarizabilities are important parameters for understanding the interaction
between photons and hadrons. For pions these quantities are poorly constrained experimentally
since they can only be measured indirectly. New experiments at CERN and Jefferson Lab are
planned that will measure the polarizabilities more precisely. Lattice QCD can be used
to compute these quantities directly in terms of quark and gluons degrees of freedom,
using the background field method. We present results for the electric polarizability for two 
different quark masses, light enough to connect to chiral perturbation theory. These are 
currently the lightest quark masses used in polarizability studies. 
\end{abstract}

\maketitle


\section{Introduction}

Hadron electromagnetic polarizabilities parametrize the response of the charge and current
distribution to an external electromagnetic field. They encode information about the structure
of the hadron. More specifically, to the lowest orders in the field strength the interaction 
between a hadron and an external electromagnetic field is parametrized by the effective hamiltonian:
\beq
H_\text{em}=-\bm\mu\cdot\bm B-\frac12\alpha\bm E^2-\frac12\beta \bm B^2 + \ldots \,,
\label{eq:1}
\eeq
where $\mu$ is the magnetic dipole of the hadron, $\alpha$ is its electric polarizability, 
and $\beta$ is the magnetic one. Here we will focus on the electric polarizability for the pions.

For pions chiral perturbation theory~($\chi$PT) predicts that 
at ${\cal O}(p^4)$, the leading order contribution for polarizabilities, we 
have $\alpha_\pi+\beta_\pi=0$~\cite{Donoghue:1989sq}.
Using this relation and an $SU(3)$ chiral Lagrangian, pion electric polarizabilities 
at ${\cal O}(p^4)$ are~\cite{Bijnens:1987dc}
\beqs
\alpha_{\pi^\pm} &= \frac{4\alpha}{m_\pi f_\pi^2} (L_9^r + L_{10}^r) 
\approx 2.68(42)\times 10^{-4}\fm^3 \,, \\
\alpha_{\pi^0} &= -\frac\alpha{96\pi^2 m_\pi f_\pi^2} \approx -0.50\times 10^{-4}\fm^3 \,.
\eeqs
A review of these calculations and other theoretical inputs relevant for pion polarizability 
can be found in the theory review included in the Da$\Phi$ne Physics handbook~\cite{Portoles:1994rc}.

Experimentally, for stable hadrons, the polarizabilities are determined from Compton scattering.
For pions experimental information comes
from Primakoff processes, radiative pion photoproduction and photon-photon fusion. For
charged pions the results indicate that $\alpha_{\pi^\pm}\in(+2, +6)\times 10^{-4}\fm^3$
and for neutral pions the current results point to $\alpha_{\pi^0}\in(-1.0, -0.3)\times 10^{-4}\fm^3$.
In some of these experiments experimental data is used to extract $\alpha_\pi-\beta_\pi$ and the value
of $\alpha_\pi$ is computed using the $\chi$PT prediction $\alpha_\pi+\beta_\pi=0$. 
Historically, there was some discrepancy between experimental and theoretical predictions for charged
pion polarizability, but more recent experiments seem to agree with $\chi$PT 
prediction~\cite{Adolph:2014kgj}. New experiments are planned at JLAB and COMPASS that will reduce
significantly the uncertainty for these quantities.

Given the rather uncertain experimental and theoretical situation, it is useful to have a calculation
for these polarizabilities directly in terms of the quark and gluon degrees of freedom. Lattice QCD
can be used to perform such a calculation using the background field method: hadron correlators are computed 
with and without an external field. The shift due to the external field is used to extract 
the polarizability. Such calculations are challenging since the effects of 
polarizability are small and they have to be disentangled from other systematic corrections due to
finite volume effects and other lattice artifacts. Additionally, for charged pions, the leading
order effect of the external electromagnetic field is to accelerate the hadrons and careful analysis
is required to separate the polarizability contribution.
A first calculation of this type~\cite{Detmold:2009dx} was carried out with
rather heavy quarks (corresponding to pion masses of $390\MeV$) and using rather large
background fields. Here we present our results for two lighter quark masses, corresponding to
pion masses of $306\MeV$ and $227\MeV$, using much weaker electric fields, well within the region
where the effects of the external fields are expected to be dominated by the polarizability contributions.
For charged pions, we also introduced an improved method of extracting the polarizabilities that
takes into account the finite volume corrections which are important for the lattice sizes used
in current simulations.


\section{Calculation details}

In this section we review the methods used to extract the electric polarizability from lattice QCD. 
We use the background field method~\cite{Martinelli:1982cb}. For neutral particles this amounts 
to computing the hadron mass with and without a background electric field and use Eq.~\ref{eq:1}
to extract the polarizability from the energy shift. In the continuum the electric field is introduced
by coupling it to the covariant derivative
\beq
D_\mu = \partial_\mu-i g G_\mu - i q A_\mu \,,
\eeq
where $A_\mu$ is the vector potential corresponding to the background field and $q$ is the charge
of the quark. On the lattice this amounts to a modification of the gauge links by a 
{\it em-factor} controlled by the vector potential
\beq
U'_\mu = e^{-iq a A_\mu} U_\mu \,,
\eeq 
where $a$ is the lattice spacing. One thing to note is that real electric fields lead to a real 
{\it em-factor}
on the lattice, since the lattice formulation involves Euclidean time. When the theory is analytical 
in the strength of the electric field ${\cal E}$ around ${\cal E}=0$, then we can use imaginary
values for the background electric field which lead to a $U(1)$ {\it em-factor}. The only change is that
the value of the mass shift is connected to the polarizability using a slightly modified expression.
To be more specific, if we use imaginary electric fields, one possible gauge choice is
\beq
U_1' = e^{-iaq{\cal E}t} U_1 \quad\text{and}\quad U_\mu' = U_\mu \text{ for } \mu \in \{2, 3, 4\} \,.
\eeq 
In this case the polarizability is connected to the phase shift 
via~\cite{Shintani:2006xr,Alexandru:2008sj}
\beq
\Delta m = +\frac12 \alpha {\cal E}^2 \,.
\label{eq:6}
\eeq

We mentioned that we need analyticity in order to be able to use imaginary values for the electric
field. In the infinite volume and even on finite boxes with periodic boundary conditions the theory
is not analytic around the ${\cal E}=0$ point due to the Schwinger mechanism. The instability is due
to the fact that the cost of creating a charged particle-antiparticle pair can be compensated by 
aligning the resultant dipole with the electric field. In infinite volume the dipole can be made 
arbitrarily large by separating the pair and the instability can then be triggered by arbitrarily
small external fields. This instability can be removed by limiting the maximal distance in the
direction of the field. In our simulations we use Dirichlet boundary conditions in the direction
of the field. In the other spatial directions we use periodic boundary conditions. Note that the
Dirichlet boundary conditions are only used for the valence quark fields. The gluon fields use 
periodic boundary conditions in all directions.

One consequence of using Dirichlet boundary conditions is that the hadron cannot be at rest. The
lowest momentum is controlled by the distance between the hard walls, $L$, and it is of the 
order $\pi/L$. To account correctly for this motion we compute the mass shift using the energy
shift measured for moving hadrons. Assuming the continuum dispersion relations, $E^2=m^2+p^2$, and
the fact that the energy shift is very small when compared to the mass of the hadrons we have
\beq
\Delta m = \frac E m \Delta E \,,
\eeq
where $\Delta E$ is the energy shift, $E$ is the energy of the moving hadron for zero external field,
and $m$ is the mass of the hadron which is calculated
using periodic boundary conditions. This does not account for all possible effects associated
with the hadron motion and it is important to carry out an extrapolation to infinite volume.
However, only the size in the direction of the field needs to be increased since the finite volume
effects in the transverse directions are expected to decay exponentially fast and should be
negligible for the lattice sizes used in our calculations.

For our calculations we use ensembles with two mass-degenerate sea-quark flavors. The discretization
used for both sea quarks and valence quarks is nHYP clover fermions~\cite{Hasenfratz:2007rf}. We
use two different sea quark masses, one corresponding to a pion mass of $306\MeV$ and the other 
corresponding to $227\MeV$. For each ensemble we computed the polarizability for a number of
valence quark masses. This allows us to gauge the sensitivity of the polarizability on 
the sea quark mass and the valence mass separately. The parameters for each ensemble are listed in
Table~\ref{tab:1}. These are the same ensembles we used in a study of neutron electric polarizability
where the method described here was tested~\cite{Lujan:2014kia}. Unless otherwise noted these
calculations were carried out using neutral sea quarks, with periodic boundary conditions in all
directions. We use around $20$ point sources for each configurations, and $5$ different couplings 
to the background field for each source, which require thousands of inversions for each configuration.
To compute these efficiently we use our own GPU Wilson dslash kernel 
implementation~\cite{Alexandru:2011sc} with an efficient multi-mass inverter~\cite{Alexandru:2011ee}.


\begin{table}
\begin{tabular}{lccccccccc}
\hline
\tablehead{1}{r}{b}{ensemble}  
  & \tablehead{1}{r}{b}{size}
  & \tablehead{1}{r}{b}{spacing}
  & \tablehead{1}{r}{b}{$N_\text{cfg}$}
  & \tablehead{1}{c}{b}{$m_1$}
  & \tablehead{1}{c}{b}{$m_2$}
  & \tablehead{1}{c}{b}{$m_3$}
  & \tablehead{1}{c}{b}{$m_4$}
  & \tablehead{1}{c}{b}{$m_5$}   \\
\hline
EN1 & $24^3\times48$ & $0.125(2)\fm$ & 300 & $269(1)\MeV$ & $\bm{306(1)}\MeV$ & $339(1)\MeV$ & $368(1)\MeV$ 
& ---\\
EN2 & $24^3\times64$ & $0.122(1)\fm$ & 450 & $\bm{227(2)}\MeV$ & $283(1)\MeV$ & $322(1)\MeV$ & $353(1)\MeV$ 
& $382(1)\MeV$ \\
\hline
\end{tabular}
\caption{Parameters for the ensembles used in this study. For each ensemble a number of valence
quark masses were used; we indicate here the pseudo-scalar masses as a proxy for the valence quark mass.  The values indicated with bold figures correspond to the unitary points, that is the points where the
valence and sea quark masses are equal.}
\label{tab:1}
\end{table}

\section{Neutral pions}

In this section we discuss our results for the electric polarizability of neutral pions. 
We first note that the correlator for $\pi^0$ has disconnected contributions even when the
sea quarks flavors have the same mass. The disconnected contributions only vanish if the
theory is isospin symmetric, which is not the case when the quarks are charged differently.
The disconnected contributions are very difficult to compute and are usually not included in
lattice calculations since their effect is expected to be small. For electric polarizability
the effect of excluding the disconnected contributions is expected to be more significant: 
$\chi$PT predicts that the value of the polarizability will be small and positive~\cite{Detmold:2009dx}
and relatively insensitive to the quark mass. In effect, the particle described by the lattice
calculations is similar to the pion with the flavor content $\bar{d}\gamma_5 u$ in a theory where
both $u$ and $d$ quarks have the same charge.

\begin{figure}[b]
  \includegraphics[width=.5\textwidth]{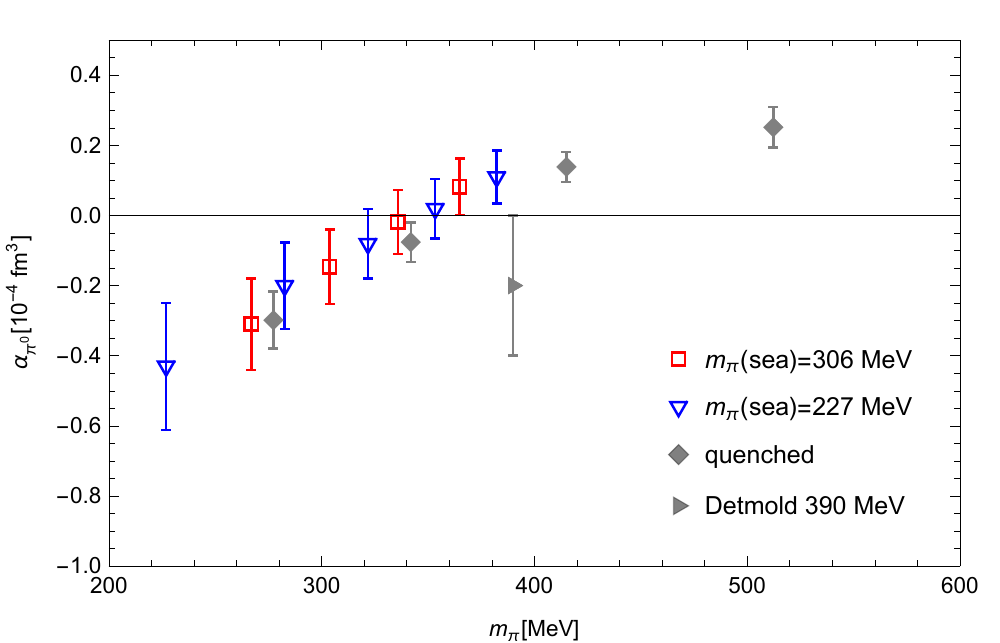}
  \includegraphics[width=.5\textwidth]{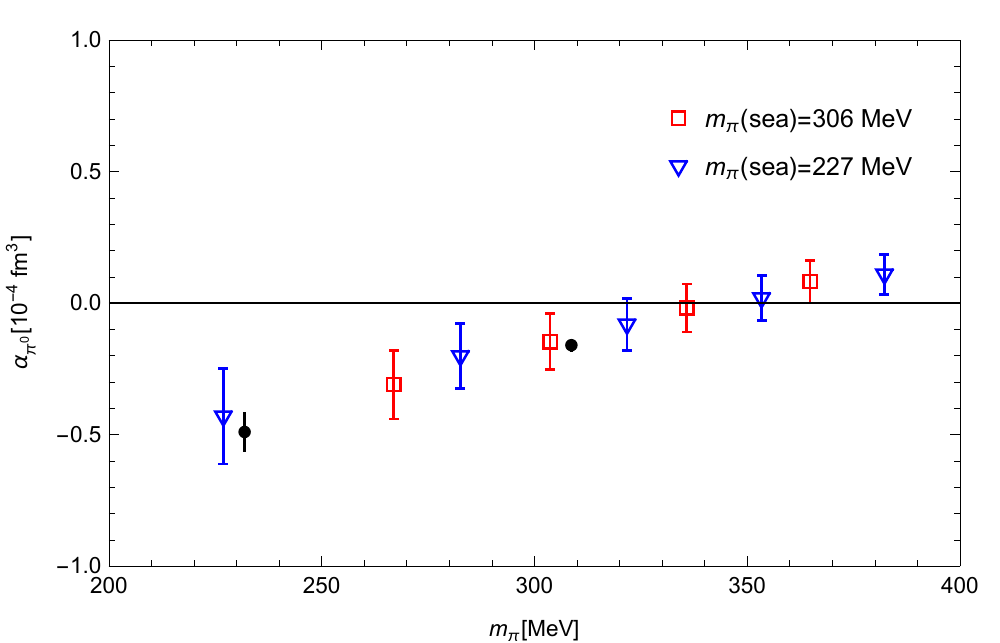}
  \caption{Neutral pion electric polarizability as a function of the valence quark mass.
  Left panel: values extracted on the ensembles EN1 and EN2.  
  The quenched data is taken from~\cite{Alexandru:2009id} and the
  $m_\pi=390\MeV$ data point is taken from~\cite{Detmold:2009dx}.
  Right panel: extrapolations
  to infinite volume (black points) compared to the finite volume values.
  }
  \label{fig:1}
\end{figure}

Our results are presented in Fig.~\ref{fig:1}. In the left panel we show the results for both
EN1 and EN2, together with the results from a previous quenched study. We see that the polarizability
is sensitive to the valence quark mass, turning negative for $m_\pi\approx350\MeV$. 
On the other hand, the polarizability
seems to be insensitive to the mass of the sea quarks, since the curves for quenched calculation
and our results for EN1 and EN2 agree with each other within the error-bars.


\begin{figure}[t]
  \includegraphics[width=.5\textwidth]{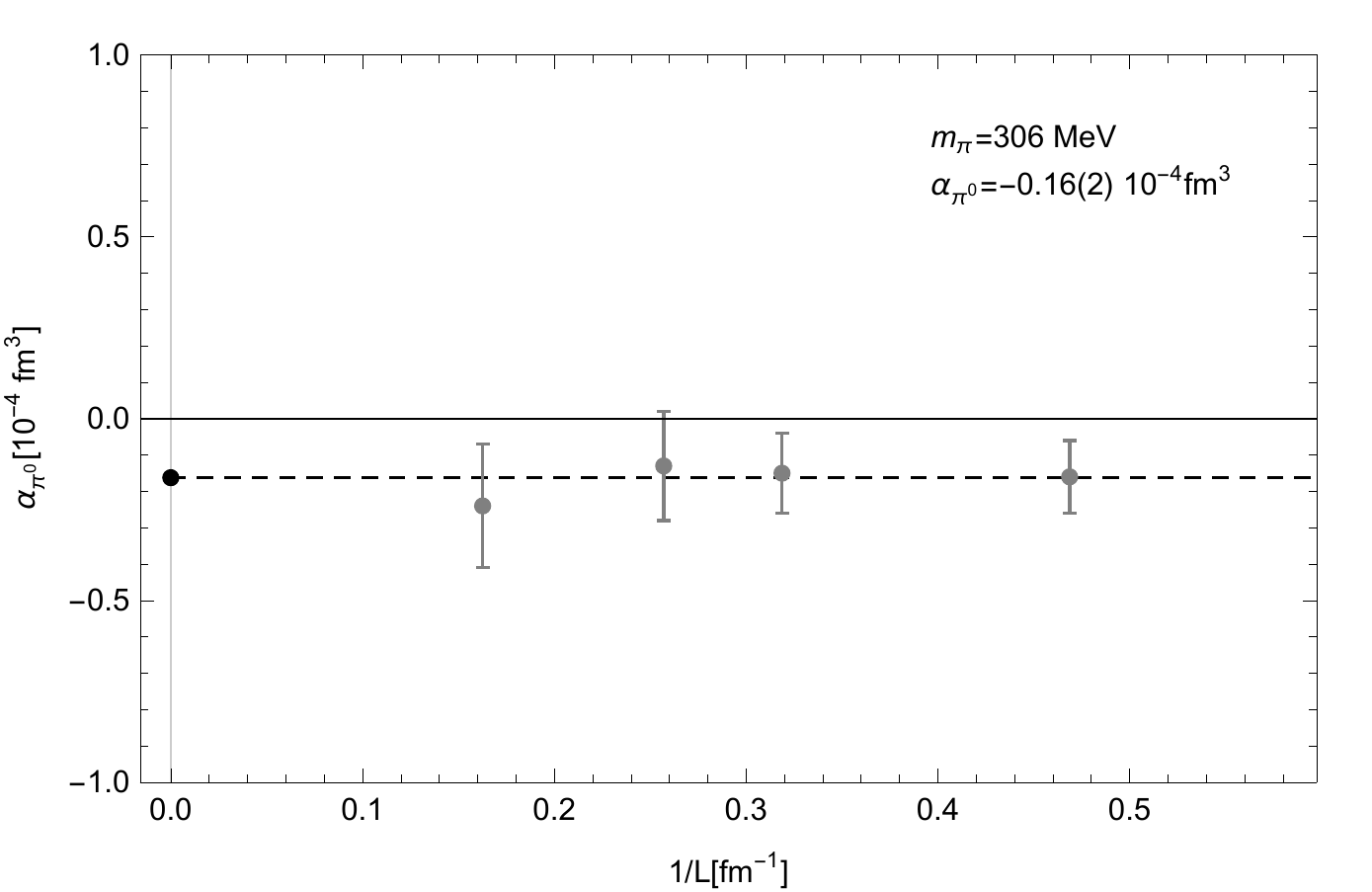}
  \includegraphics[width=.5\textwidth]{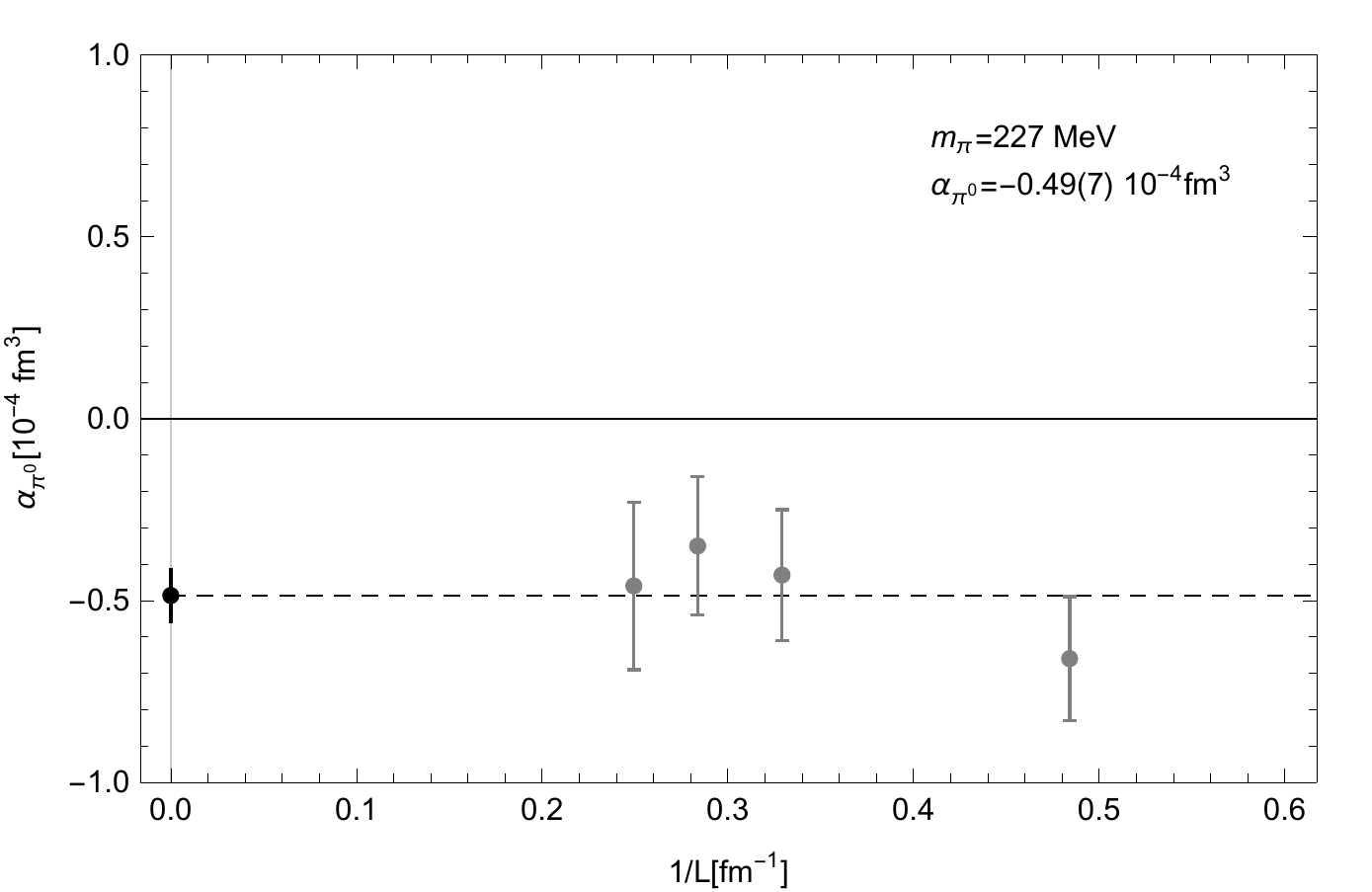}
  \caption{Neutral pion electric polarizability as a function of the size of the lattice
  in the direction of the electric field for EN1(left) and EN2(right).
  }
  \label{fig:2}
\end{figure}

These results are a bit different from the expectations from $\chi$PT, but they are 
consistent with results from previous lattice studies~\cite{Alexandru:2009id, Detmold:2009dx}.
Possible sources for the discrepancy between the lattice results and $\chi$PT expectations
are the finite volume effects and the neutral sea quarks. To study the finite volume effects,
we generated ensembles with the same parameters but with a series of increasing sizes in the
direction of the electric field~\cite{Lujan:2014qga}. For pions we found that the polarizability
depends very little on the size of the box and we can fit the results with a constant as can
be seen in Fig.~\ref{fig:2}. The results for the infinite volume extrapolation are presented
in the right panel of Fig.~\ref{fig:1}. It is clear that the negative trend remains and that
the finite volume effects are not the source of discrepancy. Charging the sea quarks is very 
challenging. A first study in this direction was carried out recently~\cite{Freeman:2014kka}
and the results seem to indicate that the polarizability might turn positive when the sea
quarks are charged. However, the error bars are large and no definite conclusion can be derived.

\section{Charged pions}

Computing the electric polarizability for charged pions presents additional challenges
since the hadrons are accelerated by the external field and their correlators are no
longer described by a sum of exponential functions. Traditional lattice spectroscopy methods
cannot be used to extract the mass of the hadrons. A solution for this problem was
proposed in~\cite{Detmold:2009dx}: compute the correlator for a charged particle in
a constant electric field and use this functional form to fit the lattice correlator.
The range of the fit has to be adjusted so that the lattice correlator is dominated by
the lowest mass particle. The functional form for the correlator was derived using
Schwinger proper time techniques using imaginary electric fields and Euclidean time.

\begin{figure}[t]
  \includegraphics[width=.5\textwidth]{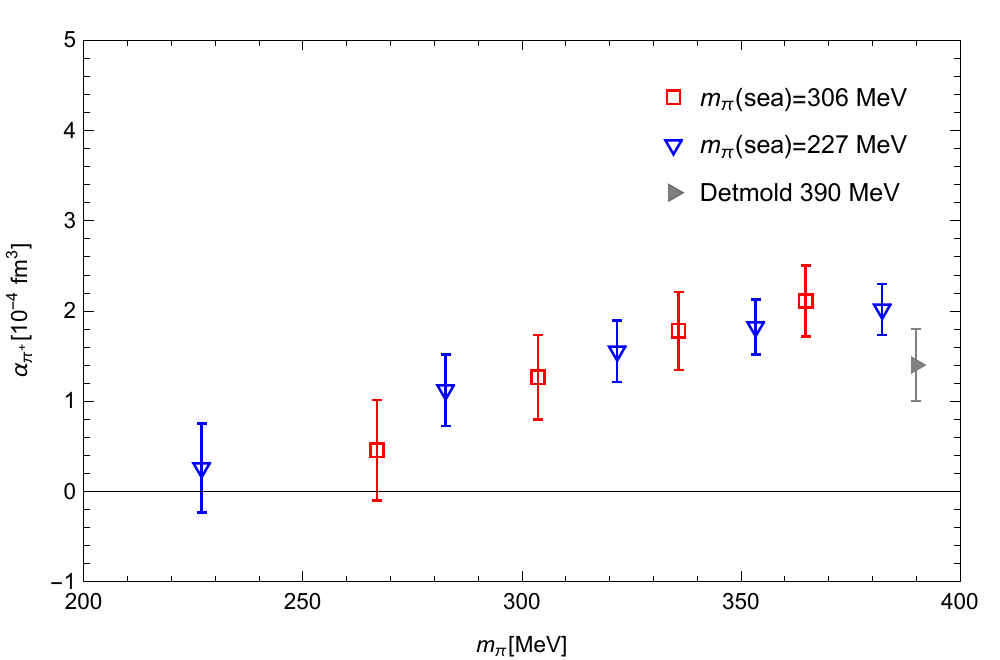}
  \includegraphics[width=0.5\textwidth]{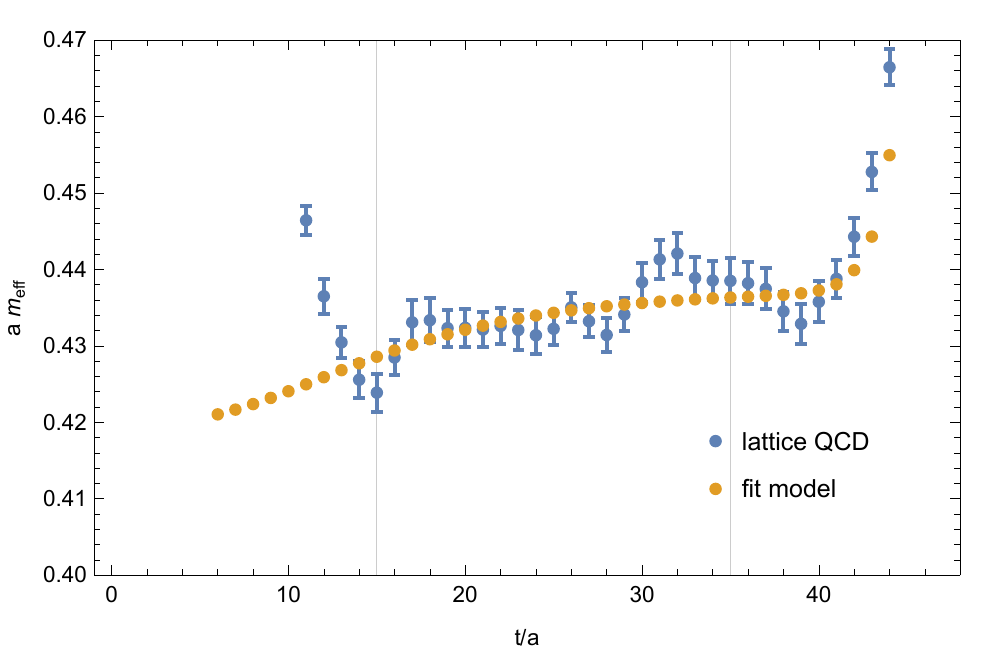}
  \caption{
  Left: electric polarizability for the charged pions as a
  function of the valence quark mass. The data for $m_\pi=390\MeV$ is taken from~\cite{Detmold:2009dx}.
  Right: effective mass for a charged pion correlator together with the
  scalar particle correlator determined from the fit. The fitting range is indicated
  by the vertical bars. 
  }
  \label{fig:4}
\end{figure}

\begin{figure}[t]
  \includegraphics[width=0.99\textwidth]{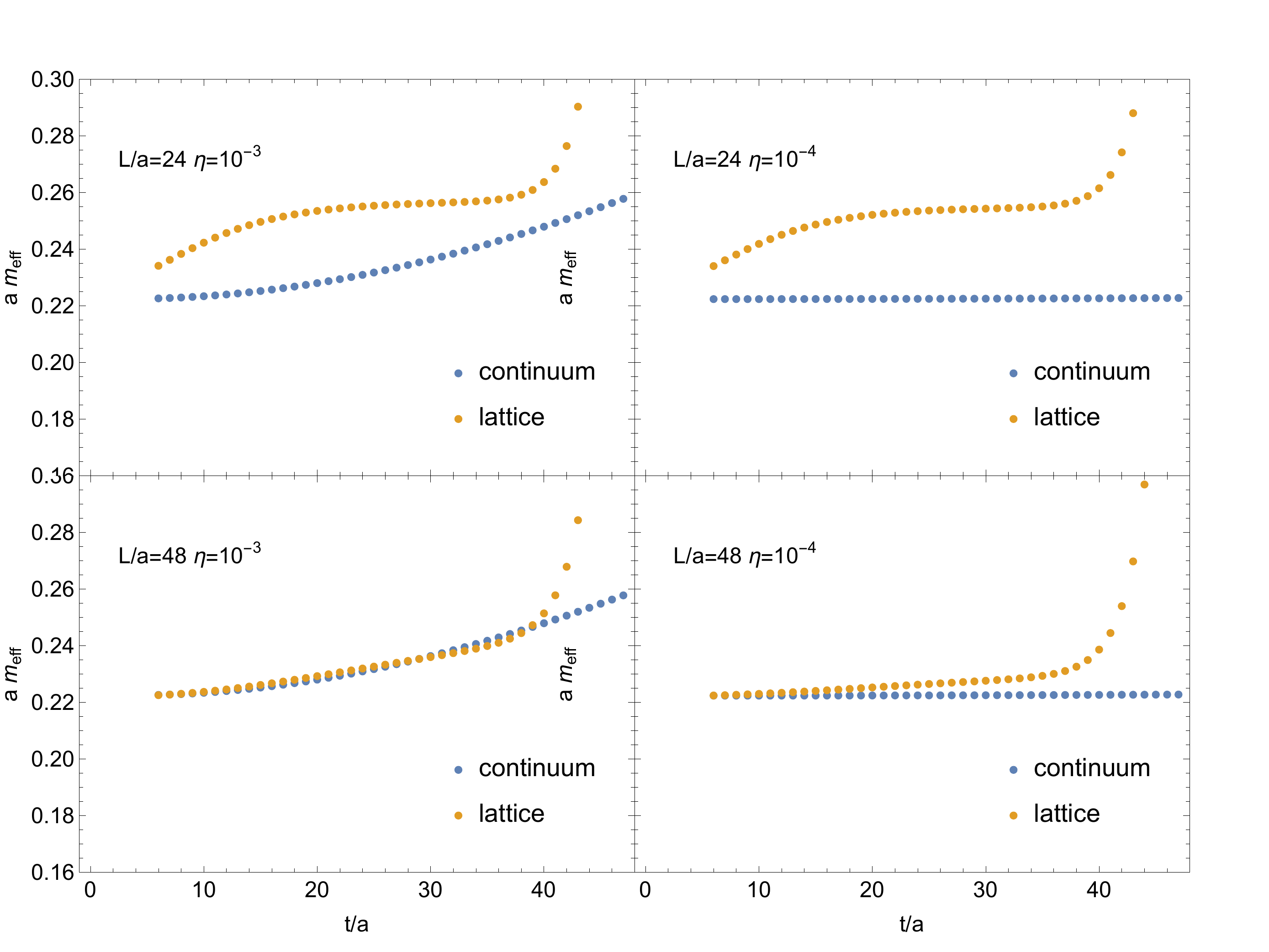}
  \caption{Effective mass for a scalar particle correlator in the presence of a background
  field. The top panels correspond to a ``small'' box and the lower ones to a larger box.
  The left panels correspond to a strong electric field $\eta\equiv q a^2 {\cal E}=10^{-3}$,
  whereas the right ones correspond to a weaker field $\eta=10^{-4}$, similar to the one
  used in our lattice study. The infinite volume result is labeled ``continuum''.
  }
  \label{fig:3}
\end{figure}

Our original attempts to fit this functional form to our correlators failed. The cause
for this was the finite volume effects. In Fig.~\ref{fig:3} we plot the
infinite-volume correlator against a free scalar correlator computed in a finite box using
lattice methods. Note that we only see an agreement when using large boxes and strong electric
fields, much stronger than the one we used in our study. To account for this discrepancy we
decided to fit our lattice QCD data to finite-volume scalar propagators computed using
numerical methods. This allows us to match the Dirichlet boundary conditions used in our
lattice QCD study with relative ease. In the right panel of Fig.~\ref{fig:4} we show the effective
mass for a lattice QCD correlator together with the fitted result using the scalar corrector.
The agreement is very good in the fit window indicated in the figure.

Using this fitting function we perform a correlated fit where both the zero and
non-zero field residues are minimized. The mass shift is then converted to polarizability
using Eq.~\ref{eq:6}. The results for our ensembles are shown in the left panel
of Fig.~\ref{fig:4}. We note again that the polarizability decreases with decreasing
valence quark mass, but it remains positive throughout the range of the masses used
in this study. The results are fairly insensitive to the sea quark mass 
(as it is the case for neutral pions), but this might
be an artifact of the fact that the sea quarks are not charged. The values are smaller
than the expectation from $\chi$PT and the discrepancy increases as we lower the quark
mass towards the physical point.

\section{Future directions}

Lattice QCD calculations of electric polarizabilities for pions are challenging.
The two basic challenges, common to all hadrons, are the fact that the finite volume
effects vanish only algebraically in the infinite volume limit and that charging the
sea quarks is numerically expensive. 
Our initial studies of these effects are very promising~\cite{Lujan:2014qga,Freeman:2014kka}. 
To reduce the error-bars, especially as we reduce the quark mass towards the physical point,
further improvements are needed. A better understanding of the finite-volume corrections 
in the presence of Dirichlet boundary conditions,
hopefully with a functional form derived from effective field theory, would greatly 
improve the efficiency of lattice calculations. Perturbative reweighting is effective in
taking into account the charge of the sea quarks, and we are currently working on improvements
that should allow us to extend this method to larger lattices to compute the infinite volume limit.

For pions new challenges arise: for neutral pions the inclusion of 
disconnected diagrams and for charged pions the acceleration due to the external field
produces non-standard correlators. The infinite volume correlator cannot be used to fit
lattice QCD results unless we use large boxes and strong fields. The method based on the 
finite-volume scalar correlator produces reasonable results, but we need to test it more thoroughly. 
We plan to compute the infinite volume limit for the electric polarizability for charged 
pions and compare it with the experimental results and $\chi$PT predictions.
We note that for charged baryons a different fitting function is needed.
The technique used in this paper can be trivially extended to compute this fitting function.


\begin{theacknowledgments}
  The calculations presented here were performed on the following GPU clusters: GWU IMPACT clusters
  and GWU  Colonial One cluster. This work is supported in part by the NSF CAREER grant PHY-1151648 
  and the U.S. Department of Energy grant DE- FG02-95ER-40907.
\end{theacknowledgments}



\bibliographystyle{aipproc}   

\bibliography{my-references}

\end{document}